\documentclass[prl,amsmath,amssymb,showpacs]{revtex4}

\begin{document}

\title{Statistics, Condensation and the Anderson-Higgs Mechanism: \\
The Worldline Path Integral View}

\author {Jian-Huang She}
\email {she@lorentz.leidenuniv.nl}
\affiliation {Instituut Lorentz voor de theoretische natuurkunde, 
Universiteit Leiden, P.O. Box 9506, NL-2300 RA Leiden, 
The Netherlands}
\author {Darius Sadri}
\email {sadri@lorentz.leidenuniv.nl}
\affiliation {Instituut Lorentz voor de theoretische natuurkunde, 
Universiteit Leiden, P.O. Box 9506, NL-2300 RA Leiden, 
The Netherlands}
\author {Jan Zaanen}
\email {jan@lorentz.leidenuniv.nl}
\affiliation {Instituut Lorentz voor de theoretische natuurkunde, 
Universiteit Leiden, P.O. Box 9506, NL-2300 RA Leiden, 
The Netherlands}

\date{\today}

\begin{abstract}
We explain, in the first quantized path integral formalism, the mechanism behind the Anderson-Higgs
effect for a gas of charged bosons in a background magnetic field, and then use the method to
prove the absence of the effect for a gas of fermions. The exchange statistics are encoded via the
inclusion of additional Grassmann coordinates in a manner that leads to a manifest
worldline supersymmetry. This extra symmetry is key in demonstrating the absence of the effect
for charged fermions.
\end{abstract}

\maketitle

\section{Introduction}

The Meissner effect \cite{Meissner}, the expulsion of magnetic fields from superconducting regions, is a salient feature of superconductivity which distinguishes it from perfect conductivity.
It can be described in a phenomenological way via the London equations \cite{Schrieffer}, but a microscopic 
understanding requires an accounting of the pairing mechanism
\cite{Cooper,Bardeen:1957mv,Bardeen:1957kj}
leading to condensation
in the ground state, and the concomitant generation of an effective mass for the photon,
\cite{Anderson:1958pb,Anderson:1963pc}.
The modern viewpoint takes the spontaneous breaking of a gauge invariance as the central idea, though of
course this is strictly speaking not correct, as a gauge symmetry can never be broken, but rather
serves as a good description in a perturbative expansion around the breaking of a global symmetry.
The breaking of a global symmetry is also relevant to the study of Bose-Einstein condensation.
An understanding of this phenomena, in the case of strongly interacting Helium and the superfluid
transition, was advanced by the introduction of the methods of first quantized path integrals
\cite{PhysRev.90.1116.2,PhysRev.91.1291}, wherein it is understood as the proliferation of worldlines
of bosons \cite{PhysRevB.36.8343,RevModPhys.67.279}. In fact, the partition function for the worldlines
can be mapped onto a second quantized Euclidean path integral (over fields) of the Landau-Ginzburg type.

The idea of spontaneously broken gauge symmetry has been used to great advantage in high energy
physics. It had long been assumed that a renormalizable theory of massive vector bosons could not be 
gauge invariant,
until it was suggested \cite{Higgs:1964ia,Higgs:1966ev,Higgs:1964pj,Hooft:1972fi,Hooft:1971fh,Hooft:1971rn}
that a microscopic gauge invariant theory involving
massless vector bosons could still account for massive vector bosons at low energy 
(like the $W^{\pm}$ and $Z^0$ in electroweak theory), if the symmetry gauged by such modes is spontaneously broken at some scale (assumed to be around a few hundred GeV for electroweak theory).
This realization guided the construction
of the electroweak theory \cite{Weinberg:1967tq,Glashow:1970gm}, now a cornerstone of the standard model of particle physics.

In electroweak theory the symmetry breaking is driven by condensation of a bosonic Higgs field, the
search for which is one of the main motivations for recent efforts in experimental high energy
physics. Various technical issues (such as the hierarchy problem) have led to the suggestion that the
Higgs particle is not in fact elementary, but gives an effective description of some as yet unknown
underlying physics (such as technicolor \cite{Farhi:1980xs}),
in much the same way that superconductivity is often described as Bose-Einstein condensation of bound
Cooper pairs \cite{Leggett}.

One might wonder if this mechanism is specific to bosons, or whether it can be realized
using fermionic constituents.
In relativistically invariant systems, condensation of fermionic operators would lead to a breaking of
Lorentz invariance.
More generaly such a condensation leads to a vacuum expectation value for the fields of the form
$\langle0|\hat{O}|0\rangle=v$, with $|0\rangle$ the
vacuum state of the system, and $\hat{O}$ either a bosonic or fermionic operator. 
Since fermionic operators connect bosonic states to fermionic ones and vice-versa, and it assumed that the vacuum state
has a definite character (in fact nearly always assumed bosonic), the vev $v$ must necessarily vanish.
This argument shows that spontaneous symmetry breaking driven by fermions (if at all possible)
must take a form different from the familiar picture described in terms of bosonic order parameters.
In fact this argument can also be made in the sense of superselection
rules, which limit the allowed possible observations made on a quantum system by disallowing
matrix elements between certain classes of states, and separating the Hilbert space into
superselction sectors from which linear combinations of basis vectors can not be made.
It has been suggested
\cite{PhysRev.88.101} that a superselection rule exists which obstructs the assembling of
states which are superpositions of bosons and fermions. Since a coherent state of fermions would
necessarily mix both bosonic and fermionic statistics, it is then not possible to construct
condensates of fermions. In fact, the question of whether such a superselection rule is
operative is one to be determined by experiment, and it has recently been proposed
\cite{PhysRevB.68.155419}
that observation of coherent superpositions of even and odd numbers of fermions in mesoscopic
quantum dots can be used as a test of supersymmetry.

Considering the importance of, and the many mysterious issues surrounding the mechanism of spontaneous symmetry breaking, it is valuable to have an alternative 
view of it. Here we will explore the formalism due to Feynmann \cite{PhysRev.90.1116.2,PhysRev.91.1291}, where one considers a representation 
in terms of the worldline path integral. The indistinguishability of the bosons 
translates into the recipie that one has to trace over all possible ways the worldlines can wind around
the periodic imaginary time axis. At the temperature where the average of the topological winding 
number $w$ becomes  macroscopic, $\lim_{N \rightarrow \infty} \langle w \rangle / N \neq 0$, 
the system undergoes a
phase transition either to a Bose-Einstein condensate, or a superfluid. Bose-Einstein condensation means that a
macroscopic number of particles `share the same worldline' with difference between BEC
and superfluidity being that in the latter this condensate is somewhat depleted.
This formalism turns out to be very efficient 
for numerical calculation of properties of strongly interacting bosonic systems such as Helium$^4$ \cite{PhysRevB.36.8343,RevModPhys.67.279}, where it is also shown that the average winding number corresponds directly to the superfluid density.

It is more difficult but perhaps even more interesting to consider the fermionic particles in this formalism. 
One can easily show \cite{Kleinert} that below the Fermi temperature, worldlines with macroscopic winding number also proliferate in fermionic systems; this leads to a puzzle: the macroscopic worldlines lead to a Meissner effect, via the Anderson-Higgs mechanism, in charged bosonic systems, but surely such phenomenon can not happen in charged fermionic systems \footnote{In the BCS theory of superconductivity, pairs of fermions form Cooper pairs, which have a bosonic character, and whose condensation then leads to superconductivity.}.

It is the aim of this paper to show in the worldline formalism, in a certain limit, that particles obeying fermionic
statistics can not drive an Anderson-Higgs transition.
In the next section we begin by recalling the single particle path integral for a spinless
boson, which we couple to a background magnetic field, and write the partition function for the
many-body system, from which we
compute the second order perturbative correction to the effective action.
Focusing on a special subclass of winding modes, we demonstrate the appearance of a mass for
the magnetic field.
We then  generalize this logic to the case of a spin-$1/2$ particle by way of introduction of
appropriate terms in the action for Grassmannian degrees of freedom coupled to the particle
worldlines. Underlying our observation on the behaviour of fermionic systems in this
language is the existence of a worldline (though not target space, where the particle trajectory
is embedded) supersymmetry.
The inclusion of the particle statistics leads to an additional term in the effective action,
and this addition is shown to lead to the disappearance of the effect manifested for charged bosons.

\section{Spinless Bosons in background Magnetic Field}

We begin by considering a single spinless boson, in the non-relativistic limit, whose action
reads \footnote{As shown by Feynman \cite{PhysRev.90.1116.2,PhysRev.91.1291}, interactions in the worldline 
formalism are best handled by working in a relativistic formalism, where worldlines are parameterized in 
terms of a local proper-time coordinate. Our approach, while non-relativistic, uses a similar parametrization 
of worldlines.
}
\begin{equation} \label{free-rel-action}
  {\cal A}_{e,0} \:=\: \int_{\tau_a}^{\tau_b} d\tau \: \frac{M}{2}{\dot x}^2(\tau) \ ,
\end{equation}
with $\tau$ the proper time along the particle's worldline.
In the presence of the electromagnetic field, one needs to add the interaction term
\footnote{We work throughout
in units where $\hbar=c=1$.}
\begin{equation}
 {\cal A}_{e,{\rm int}} \:=\: i \: e \: \int_{\tau_a}^{\tau_b} d\tau \:
  {\dot x^i}(\tau) \: A_i(x(\tau)) \ ,
\end{equation}
where the dot in $\dot x$ denotes a derivative with respect to proper time of the particle,
which should not be confused with the Euclidean time in target space. Here $i,j=1,\cdots,d$, with $d$ the dimension of space.
We shall only be interested in the study of particles immersed in an external magnetic field.
Hence, in the following we set the electric field to zero, $E^i=0$, and consider only the response to a magnetic field $B^i$. We drop the inter-particle Coulomb repulsion.

Since we are interested in using the single- and many-body path integrals in first-quantized form,
we are restricted to considering non-relativistic physics. Standard problems with negative probabilities
and pair production would force us to rely on the second-quantized quantum field theory language to
address the relativistic problem.

We now study the condensation of bosonic particles in a background magnetic field,
giving a new vantage point on the Meissner effect, before we turn to apply the same ideas to the study
of fermionic systems.
The partition function of $N$ identical bosons sums over all permutations $\cal{P}$ of the particle coordinates (with no relative minus sign)
\begin{equation}
  Z_N=\frac{1}{N!}\int dx_1\cdots\int dx_N\sum_{\cal P}\prod_i
  \left( x_{p(i)},\beta|x_i,0\right) \ ,
\end{equation}
with
\begin{equation}
  \left( x_{p(i)},\beta|x_i,0\right) \:\equiv\:
  \int_{x_i}^{x_{p(i)}} {\cal D} x \:
  e^{-{\cal A}_e^{(i)}} \ .
\end{equation}
We study the system at finite temperature, which is reflected in the fact that the worldlines wrap around the imaginary (thermal) time direction, with $\tau$ running from $0$ to $\beta$, i.e. the action
involves $\mathcal{A}=\int_0^\beta \dots d\tau$.

Consider a general partition of the orbits of $N$ particles 
grouped into different winding cycles via permutation,
\begin{equation}
  N=\sum_{w=1}^N w \ C_w \ .
\end{equation}
In this decomposition
we keep track of the number of cycles (which we denote by $C_w$) each of length $w$, so that
with each permutation we associate a series of numbers $C_w$, with $w=1,...,N$.
Then, a sum over all permutations can be rewritten as a sum over all integers assigned to
the various $C_w$, subject of course to an overall constraint, this constraint being that the total
length of all cycles taken together must be $N$ (for a discussion of this point, see \footnote{See appendix of \cite{zaanen-2008}.}).
The number of permutations with such a decomposition is
\begin{equation}
  M(C_1,C_2,\cdots,C_N)=\frac{N!}{\prod_{w=1}^N{C_w!}w^{C_w}} \ .
\end{equation}
The partition function of $N$ bosons is a summation over different partitions
\begin{equation}
  Z^{(N)}(\beta)=\frac{1}{N!}
  \sum_{\left\lbrace C_1,\cdots,C_N\right\rbrace }M(C_1,\cdots,C_N)\prod_{w=1}^N[Z(w\beta)]^{C_w}
  \ ,
\end{equation}
where for each loop one has
\begin{equation}
Z(w\beta)=\int d^dx\int_{{\cal C}_{w}} {\cal D}  x \:
  e^{-\int _0^{w\beta} d\tau \left(  \frac{M}{2}{\dot x}^2
  + i\frac{e}{c}{\dot x}_i A^i\right)} \ ,
\end{equation}
with the loop winding $w$ times around the imaginary time direction.

We consider first a single winding loop with length $w$.
To study the Meissner effects, we employ a standard procedure, namely to first expand the interaction part of the partition function $\exp(-ie\int d\tau {\dot x}_i A^i)$ as a power series,
and then compute the average of each term with respect to the free particle action, which leads to corrections of the form $\left\langle A^n\right\rangle_0$, i.e. averages taken with respect to the free system.
Define for this particular winding loop the correction to the effective action to be
$\Delta \Gamma(w\beta)=Z(w\beta)-Z_0(w\beta)$. Here the lowest order non-trivial term is of order $A^2$, and its contribution to the Euclidean effective action reads
\begin{equation}
  \Delta \Gamma(w\beta) \:=\:
  \frac{e^2}{2} \Big< \int d\tau_1 \int d\tau_2 \
  {\dot x}_i(\tau_1) A^i(x(\tau_1)) \
  {\dot x}_j(\tau_2) A^j(x(\tau_2))
  \Big>_0 \ ,
\end{equation}
where by definition the average of the operator $O$ with respect to the free action is
$<O[x]>_0 \sim \int d^dx \int \mathcal{D}x e^{-A_0} O[x]$, up to a normalization factor.
We will work with the Fourier transform of the gauge potential
$A(x)=\int\frac{d^dk}{(2\pi)^d}e^{ikx}\tilde{A}(k)$, and will have to evaluate expectation values of
the form $<e^{ik_1 x(\tau_1)} e^{ik_2 x(\tau_2)} \dot{x}(\tau_1) \dot{x}(\tau_2)>_0$.
To do so, we expand the position as the sum of an average and a fluctuation part
$x(\tau)=x_0+\delta x(\tau)$, where the average is the same for all coordinates appearing above.
The desired expectation value then factorizes into
$<e^{i(k_1+k_2) x_0}>_0 <\delta\dot{x}(\tau_1) e^{ik_1 \delta x(\tau_1)}
\delta\dot{x}(\tau_2) e^{ik_2 \delta x(\tau_2)}>_0$
(indices have been suppressed in an obvious fashion).
The first factor is easily shown to result in a delta function
$(2\pi/L)^d\delta(k_1+k_2)$,
ensuring momentum conservation,
and we evaluate the second factor by applying Wick's theorem.
We get that
\begin{equation}
\label{gam}
  \Delta \Gamma(w\beta) = \frac{e^2}{2 L^d}
  \prod_{\alpha=1}^2 \int d^dk_{\alpha} \ \delta(k_1+k_2) \
  \tilde{A}^i(k_1) \tilde{A}^j(k_2)
  \int_0^{w\beta} \!\!\!\! d\tau_1\int_0^{w\beta} \!\!\!\! d\tau_2
  \left[\frac{\partial^2G}{\partial\tau_1\partial
  \tau_2}\delta^{ij}+k_1^{i}k_1^{j}\frac{\partial G}{\partial\tau_1}\frac{\partial G}{\partial\tau_2}\right]
  e^{(k_1^2+k_2^2) G^{\prime}} \ .
\end{equation}
Note that the $\tau$ integrals now run from $0$ to $w \beta$.
Here we used the standard language of Green's functions, which is explained as follows:
for a single particle, the Green's function is defined as 
\begin{equation}
  \delta^{ij}G_1(\tau_1,\tau_2)\equiv\left\langle x^i(\tau_1)x^j(\tau_2) \right\rangle_0 \ , 
\end{equation}
which in the path integral formalism reads
\begin{equation}
  G_1(\tau_1,\tau_2)=\int d^dx\int{\cal D}x \: e^{-{\cal A}_{e,0}}x(\tau_1)x(\tau_2) \ .
\end{equation}
This Green�s function can be derived from the zero frequency limit of the finite temperature
harmonic oscillator, after a subtraction of an infinite contribution due to the zero Matsubara frequency,
yielding
\begin{equation}
\label{greens-function-1}
  G_1(\tau_1,\tau_2)=-\frac{\tau_1-\tau_2}{2}+\frac{(\tau_1-\tau_2)^2}{2\beta}+\frac{\beta}{12} \ .
\end{equation}
For a many particle system, we define the Green's function for a particular permutation pattern as
\begin{equation}
  G_N(\tau_1,\tau_2) \equiv
  \int dx_1 \cdots dx_N
  \int_{x_1}^{x_{p(1)}} {\cal D} x^{(1)} \cdots \int_{x_N}^{x_{p(N)}} {\cal D} x^{(N)} \
  e^{- \left({\cal A}_e^{(1)}+\cdots+{\cal A}_e^{(N)}\right)} x(\tau_1)x(\tau_2) \ ,
\end{equation}
or
\begin{equation}
  \delta^{ij}G_N(\tau_1,\tau_2) \:\equiv\:
  \left\langle x^i(\tau_1)x^j(\tau_2) \right\rangle_{\cal P} \ ,
\end{equation}
and the result for the chosen winding loop is just the one-particle Green's function with $\beta$ replaced by $w\beta$ 
\begin{equation}
  G_w(\tau_1,\tau_2)=-\frac{\tau_1-\tau_2}{2}+\frac{(\tau_1-\tau_2)^2}{2w\beta}+\frac{w\beta}{12} \ .
\end{equation}
Furthermore, the subtracted Green's function \footnote{This Green's function can be related to the zero frequency limit
of a harmonic oscillator Green's function, and the subtraction removes a divergent term arising in this
limit; the divergence can be traced to the contribution of the zero Matsubara frequency.}
is defined as
\begin{equation}
\label{greens-function-2}
G^{\prime}(\tau_i,\tau_j)\equiv G(\tau_i,\tau_j)-G(\tau_i,\tau_i).
\end{equation}
We now proceed to calculate the correction to the effective action \eqref{gam}.
The $\delta$-function forces $k_1=-k_2$, and since the integrand only depends on the difference
$\tau_1-\tau_2$, one of the $\tau$ integrals can be easily calculated, giving only an overall factor. In this way
\eqref{gam} simplifies to  
\begin{equation}
\label{mast} 
  \Delta\Gamma \:=\: \frac{w\beta e^2}{2M^2 L^d} \int d^dk \
  \tilde{A}^i(k) \tilde{A}^j(-k) \ \Omega_{ij}(k) \ ,
\end{equation}
where all the relevant information is encapsulated in the momentum dependent function
\begin{equation}
\label{ome} 
   \Omega_{ij}(k) = (k^2\delta^{ij}-k^i k^j)
\int_0^{w\beta}  d\tau \left( -\frac{1}{2}+\frac{\tau}{w\beta} \right)^2 
  e^{\frac{k^2}{M}(-\frac{\tau}{2}+\frac{\tau^2}{2w\beta})} \ .
\end{equation}

When $w$ is finite, including the case with only a single particle, the $k^2\delta^{ij}-k^i k^j$ term will give rise to two differentials on the gauge field when transformed back to real space, giving the spatial part of the well known vacuum polarization
\begin{equation}
  \int d^{d} x \ F_{ij}(x) \ \Pi(-\partial^2) \ F^{ij}(x) \ ,
\end{equation}
with the field strength $F_{ij}(x)=\partial_i A_j(x)-\partial_j A_i(x)$. $\Pi(-\partial^2)$ is the self-energy of the electromagnetic field, with corrections arising from polarization effects induced by the bosons
which are coupled to the electromagnetic field.

In the limit $w\to\infty$, a partial integration on Eq.(\ref{ome}) leads to the result
\begin{equation}
\Omega_{ij}(k)= (k^2\delta^{ij}-k^i k^j) \frac{M}{k^2} \left(  1-\int _{-\frac{1}{2}}^{\frac{1}{2}} dy e^{w\beta\frac{k^2}{2M}(y^2-1/4)} \right) ,
\label{omass} 
\end{equation}
with $y=\tau/w\beta-1/2$.
The second term in the bracket vanishes when $w\to\infty$. Thus the $k^2$ term is killed, and we get a mass term for the transverse component of the gauge field 
\begin{equation}
  \Delta\Gamma \:=\: \frac{m^2}{2} \int d^{d} x \ A^\perp_i A^\perp_i,
\end{equation}
with $A^\perp_i=(\delta_{ij}-\partial_i\partial_j/\partial^2)A_j$.
This is exactly the desired Meissner effect. The contribution to the mass term coming from a single winding loop is
\begin{equation}
  \Delta\Gamma(w\beta) \:\propto\: \frac{e^2n}{M}\frac{w\beta}{N}A_\perp^2,
\end{equation}
where $n=N/L^d$ is the number density. In the following, we ignore the backreaction of
$A_\perp$ on the condensate.

To get the mass of the gauge field, one needs to sum over different cycle decompositions. The correction to the effective action of the whole system is
\begin{equation} \label{sqmass} 
  \Delta\Gamma=\frac{1}{N!}\sum_{\left\lbrace C_w\right\rbrace }M({\left\lbrace C_w\right\rbrace})  
  \prod_{w=1}^N[Z_0(w\beta)]^{C_w}\sum_{w=1}^NC_w \Delta\Gamma(w\beta)
  \ .
\end{equation}
The mass term is gotten by taking the thermodynamic limit of the above equation and keeping only terms with infinite winding. Here we need to be careful about the order of limits to take. As shown above, only those permutation patterns containing infinitely long winding loops (in the thermodynamic limit), will contribute to the mass term. So we will first take the limit that the winding number goes to infinity. To do so, we also need to take the total number of particles $N$, and the size of the system $L$, to infinity, while keeping the particle number density $n=N/L^d$ fixed.  We employ a cutoff $N_c$ for the winding number, which goes to infinity as the particle number $N\to\infty$, and count only those winding loops longer than $N_c$. For example, one can take $N_c$ to be $N^\alpha$ with $0<\alpha<1$. 
In a box with side length $L$, the partition function of free bosons
for a winding loop with length $w$ is
\begin{equation}
  Z_0(w\beta)=\left( \frac{L}{\lambda\sqrt{w}}\sum_{n=-\infty}^{\infty}
  e^{-n^2(\frac{L}{\lambda})^2\frac{\pi}{w}}\right) ^d,
\label{partf} 
\end{equation}
with the thermal de Broglie wavelength $\lambda=\sqrt{2\pi\beta/M}$. Consider the case of three
dimensions, where Bose-Einstein condensation is known to occur at finite temperature, and where
$2/3<\alpha<1$. In the limit $w\to\infty$, $L/\lambda\sqrt{w}$ goes to zero and $Z_0(w\beta)\to1$ for $w>N_c$.
Thus for a particular cycle decomposition, the contribution to the mass term from the long loops reads
\begin{equation}
  \sum_{w=N_c}^NC_w \Delta\Gamma(w\beta)\sim \frac{e^2n}{MN}
  \left(\sum_{w=N_c}^N w \ C_w \right) \ A^2_{\perp}
  \ .
\end{equation}
Here $\sum_{w=N_c}^N w C_w$ just counts the number of long loops in this cycle decomposition,
and since it is only these longs loops which contribute,
the mass square becomes 
\begin{equation}
  m^2 \:=\: \frac{e^2n}{2M} \sum_{\left\lbrace C_w\right\rbrace }\frac{M({\left\lbrace C_w\right\rbrace})}{N!}
  \left( \prod_{w=1}^N[Z_0(w\beta)]^{C_w} \right)
  \ \left(\frac{\sum_{w^\prime=N_c}^N w^\prime C_{w^\prime}}{N} \right)
  \ .
\end{equation}
Since for $w$ large, $Z_0(w\beta)\simeq 1$, the temperature dependence is fully encoded in the 
contributions from small $w$.
When tempereture goes to zero, $Z_0(w\beta)\to1$ even for small $w$. The mass square thus reads 
\begin{equation}
m^2 \:=\: \frac{e^2n}{2M} \sum_{\left\lbrace C_w\right\rbrace }\frac{M({\left\lbrace C_w\right\rbrace})}{N!}\frac{\sum_{w=N_c}^NwC_w}{N}.
\end{equation} 
The combinatorial factor in the above equation can be calculated by using random permutation theory
\cite{Goncharov1,Goncharov2}. It can be rewritten in the form
\begin{equation}
  \frac{1}{N}\sum_{w=N_c}^N\sum_{k=1}^{[N/w]}kwP(C_w=k)
  \ ,
\end{equation}
where $P(C_w=k)$ is the probability to have k cycles of length $w$, and according to
\cite{Goncharov1,Goncharov2}, is 
\begin{equation}
P(C_w=k)=\frac{w^{-k}}{k!}\sum_{j=0}^{[N/w]-k}(-1)^j\frac{w^{-j}}{j!} \ .
\end{equation}
We can estimate the magnitude of the combinatorial factor as follows. For large winding number $w$, the probability to have large number $k$ of them is extremely small. Thus we can concentrate on small $k$, where $P(C_w=k)$ is approximately $\frac{1}{k!}e^{-1/w}w^{-k}$. Summing over $k$ gives roughly $\sum_kkwP(C_w=k)\simeq 1$. Taking $N_c$ to be of order $\sqrt{N}$, the combinatorial factor is then approximately one. Thus as the temperature goes to zero, the mass square goes over to
\begin{equation}
m^2 \:=\: \frac{e^2n}{M} \ .
\end{equation}
With the closed-form formula given above, one can also calculate the combinatorial factor numerically and it converges to one very quickly \footnote{For $N=100, 10000, 40000$, it is correspondingly $0.91, 0.9901, 0,995025$.}.
The bottom line is that a finite value of mass can be gotten by summing over the long winding loops.

It is conceptually the same, but technically even easier to work in the grand-canonical ensemble, where the partition function for free bosons is 
\begin{equation}
\label{free}
  F_G(\beta)= - \frac{1}{\beta} \sum_{w=1}^{\infty}
  \left( \pm 1 \right)^{w-1}
  \frac{ Z_0 (w \beta)} {w} e^{w\beta \mu} \ ,
\end{equation}
with the plus sign for bosons and the minus sign for fermions.
In the presence of an electromagnetic field, one needs only to replace $Z_0(w\beta)$ by $Z(w\beta)$.
The change in the free energy due to the background field is
\begin{equation}
\Delta F=\frac{1}{\beta}\frac{e^2n}{M} A_\perp^2\frac{1}{N}\sum_{w=N_c}^{\infty}\frac{e^{w\beta \mu}}{w}w\beta.
\end{equation}
The quantity $N_L=\sum_{w=N_c}^{\infty}e^{w\beta \mu}$ is just the number of particles residing in the long loops, or equivalently the number of particles in the condensate. This can be shown by counting the number of particles
\begin{equation}
N= \sum_{w=1}^{\infty}  Z_0 (w \beta) e^{w\beta \mu} \ .
\end{equation} 
For small winding, $Z_0(w\beta)$ is approximately $(L/\lambda\sqrt{w})^d$, while for large winding approximately $(L/\lambda\sqrt{w})^d+1$. Thus $N$ can be rewritten as 
\begin{equation}
N= \sum_{w=1}^{\infty} \left( \frac{L}{\lambda\sqrt{w}}\right)^d  e^{w\beta \mu} +\sum_{w=N_c}^{\infty}e^{w\beta \mu}\ ,
\end{equation} 
where the first term represents the number of particles $N_S$ living in the short loops, and the second term the long loops. Consider again the case of three dimensions, where the critical temperature $T_c$ is determined by setting the chemical potential $\mu=0$ and equating $N=N_S$, that is $N=\sum_{w=1}^{\infty}(L/\lambda_c)^3w^{-3/2}$, where $\lambda_c=\sqrt{2\pi/T_cM}$. In this way the ratio of the size of the system and the thermal de Broglie wavelength can be expressed as 
\begin{equation}
\frac{L}{\lambda}=\left( \frac{N}{\zeta(3/2)}\right) ^{1/3}\left( \frac{T}{T_c}\right) ^{1/2} \ .
\end{equation}
This result can be derived via standard statistical mechanics methods; see for example \cite{Kleinert} and references therein.
One can show that when Bose-Einstein condensation occurs (and thus $\mu=0$), the number of particles winding
in the short loops is 
\begin{equation}
N_S=N\left(\frac{T}{T_c} \right) ^{3/2}.
\end{equation}
Thus the fraction of the particles living in the long loops is 
\begin{equation}
\frac{N_L}{N}=1-\left(\frac{T}{T_c} \right) ^{3/2}.
\end{equation}
When there is no condensation (and thus $\mu<0$), $N_L/N$ vanishes in an obvious way. The conclusion is that the mass square of the transverse photons is determined by the number density of the condensed particles
\begin{equation}
m^2= \frac{e^2n_{\text{cod}}}{M} \ .
\end{equation}

It has been shown \cite{PhysRev.100.463} using perturbation theory that for an
ideal charged Bose gas, with the Coulomb interaction ignored and only the
magnetic coupling ${\vec p}\cdot{\vec A}$, when there is condensation, there is a Meissner
effect. The inverse screening length squared is known to be $1/\lambda^2=(e^2/M)n_{\text{cod}}$.
The calculation above agrees perfectly with this result.
Inclusion of the Coulomb repulsion between the charged particles would lead to a renormalization
of the superfluid density.
These phenomena are well understood in quantum field theory.
The above first quantized formalism gives a new picture of these well known effects.

\section{Inclusion of Spin and Fermionic Statistics}

We begin by recalling that the Hilbert space structure of a system is given by the path integral
for zero Hamiltonian ($H=0$), wherein the exponent appearing in the path integral sum consists simply of a
Berry phase term of the form $p \dot{q}$, arising from overlaps of complete sets of
position and momentum states at neighboring time slices.
Consider a non-relativistic point particle system in three space dimensions, given by the
following path integral for Grassmann variables $\theta^i(t)$ \cite{Kleinert}
\begin{equation} \label{spin-action}
  Z \:=\: \int \mathcal{D} \theta \:
  e^{i \int dt \frac{i}{4} \theta_j \dot{\theta}^j}
  \ \  \text{for} \ j=1,2,3,
\end{equation}
which is a pure Berry phase term.
This path integral takes as its starting point the classical mechanics of spin, and can be constructed
via spin coherent states \cite{Berezin:1976eg}\footnote{See
\cite{Schubert:2001he} for a general review of applications of worldline formalism to
perturbative quantum field theory}.
The momentum conjugate to $\theta^j(t)$ is given by $p_j=-(i/4)\theta^j$,
the sign arising from the Grassmann nature of $\theta$.
The equation of motion forces the variable $\theta$ to be time independent, $\dot{\theta}^j(t)=0$.
That the momentum is proportional to the position is a reflection of the fact that systems that are
first order in time derivatives (like the Dirac equation) represent constrained systems.
The second-class constraints are
\begin{equation} \label{constraint}
  \chi_j \:=\: p_j \:+\: \frac{i}{4} \theta_j \:=\: 0
  \ .
\end{equation}
The origin of the constraint lies in the fact that the transformation from Lagrangian
configuration space $q,\dot{q}$, to the phase space $q,p$, is singular with a vanishing
Jacobian determinant, which means we can not invert the velocities to solve for the momenta,
and results in a Hamiltonian which is defined only on the constraint surface.
Dirac \cite{Dirac} has shown that such systems can be handled if one extends the notion of
Poisson brackets to Dirac brackets, defined as
\begin{equation}
  \{ A, B \}_D \:=\: \{A,B\} \:-\: \{A,\chi_i\} \: C^{i,j} \: \{\chi_j,B\}
  \ ,
\end{equation}
and $C^{i,j}$ are the components of the matrix inverse of $C$, whose elements are are built
from the Poisson brackets of the constraints
\begin{equation}
  C_{i,j} \:\equiv\: \{ \chi_i , \chi_j \}
  \ .
\end{equation}
Care must be taken that the Poisson brackets of Grassmann valued fields are defined as
\cite{Henneaux:1992ig}
\begin{equation}
  \{ f(\theta_i,p_j) , g(\theta_i,p_j) \} \:=\:
  - \: \Big(
  \frac{\partial f}{\partial \theta_k} \frac{\partial g}{\partial p^k} \:+\:
  \frac{\partial g}{\partial \theta_k} \frac{\partial f}{\partial p^k}
  \Big) \ ,
\end{equation}
in order to satisfy natural algebraic properties and yield a proper quantization for fermions.
With these, we see that the Dirac bracket associated to \eqref{spin-action} is
\begin{equation}
  \{ p_i , \theta_j \}_D \:=\: - \frac{1}{2} \delta_{i,j} \ .
\end{equation}
Canonical quantization proceeds by replacing the Dirac bracket with the anti-commutator (for
fermions), as $\{,\}_D \rightarrow - i \left[ , \right]_+$.
Making this substitution and enforcing the constraints \eqref{constraint}, we have the operator
equation
\begin{equation} \label{op-eqn}
  \big[ \hat{\theta}_i  \: , \: \hat{\theta}_j \big]_+ \:=\:2 \delta_{ij} \ .
\end{equation}
In three dimensions, the operators $\theta^i$ can be defined via their matrix elements as
\begin{equation} \label{matrix-element}
  \langle \alpha | \hat{\theta}^i | \beta \rangle \:\equiv\: \sigma^i_{\alpha,\beta} \ ,
\end{equation}
with the range of the spinor indices $\alpha,\beta=1,2$, and the $\sigma$ matrices being the standard
Pauli matrices, after which \eqref{op-eqn} is a simple identity for the Pauli matrices.
A classical spin vector can also be defined
\begin{equation} \label{spin-vector}
  S^i \:=\: - \frac{i}{4} \epsilon^{ijk} \theta^j \theta^k \ ,
\end{equation}
which after canonical quantization, gives an operator representation of the spin algebra
\begin{equation}
  \big[ \hat{S}^i \:,\: \hat{S}^j \big]_- \:=\: i \epsilon^{ijk} \hat{S}^k \ .
\end{equation}

Now consider the addition to the zero Hamiltonian path integral a term
describing the coupling of a spin to an external magnetic field $\vec{B}$,
$H=-\vec{S} \cdot \vec{B}$, the spin vector having already been defined by \eqref{spin-vector}.
Taking account of \eqref{matrix-element}, we see that \cite{Kleinert} matrix elements of
the operator
\begin{equation} \label{mag-field}
  e^{i \int dt \vec{B} \cdot \frac{\vec{\sigma}}{2}}
  \ ,
\end{equation}
has the path integral representation
\begin{equation} \label{spin-action-2}
  \int \mathcal{D} \theta \:
  e^{i \int dt \frac{i}{4} \left( \theta_j \dot{\theta}^j \:+\:
   \epsilon^{jkl} B^j \theta^k \theta^l \right) }
  \ .
\end{equation}
The trace of the operator \eqref{mag-field} can then be computed by summing the path integral
\eqref{spin-action-2} over all anti-periodic paths, for which
$\theta^j(\tau_b)=-\theta^j(\tau_a)$. For zero external magnetic field this fixes the
normalization of the Berry phase term \eqref{spin-action}. This normalization simply counts the
dimension of the relevant spinor representation in $d$ dimensions
\begin{equation} \label{norm}
  \int \mathcal{D} \theta \exp\left[ {\frac{i}{4}
  \int_{\tau_a}^{\tau_b=\tau_a}d \tau \: \theta_j \dot{\theta}^j} \right] \:=\: 2^{d-2}
  \ ,
\end{equation}
which for $d=3$ coincides with the dimension of the Pauli matrices \eqref{matrix-element}.

The free particle action \eqref{free-rel-action} can now be modified for the inclusion of
spin degrees of freedom as follows \cite{Kleinert}
\begin{equation} \label{susy-free}
  {\cal A}_{e,0}=\int _{\tau_a}^{\tau_b}d\tau \left( \frac{M}{2}{\dot x}^2(\tau)-
  \frac{i}{4}\theta_j(\tau){\dot \theta}^j(\tau)\right) \ ,
\end{equation}
while the coupling to the electromagnetic field becomes
\begin{equation} \label{susy-int}
  {\cal A}_{e,{\rm int}} \:=\: i \frac{e}{c} \: \int_{\tau_a}^{\tau_b} d\tau
  \left( {\dot x}_j A^j+i\frac{1}{4M}F_{jk}\theta^j\theta^k\right) \ .
\end{equation}
This action contains an ``orbital" contribution associated with the particle's motion $x$,
together with a ``spin" contribution arising from the Grassmann coordinates $\theta$.
The Grassmann field obeys antiperiodic boundary condition with $\theta(\tau_b)=-\theta(\tau_a)$, in contrast to the periodic boundary condition for $x$.


An important property of the spinfull interacting action
\eqref{susy-free} and \eqref{susy-int},
is an underlying
worldline supersymmetry, mixing the bosonic and fermionic degrees of freedom, given by
\cite{Brink:1976uf,Kleinert}
\begin{eqnarray} \label{susy-trans}
  \delta x^j(\tau) &=& i \alpha \theta^j(\tau) \ , \nonumber\\
  \delta\theta^j(\tau) &=& \alpha{\dot x}^j(\tau) \ ,
\end{eqnarray}
with $\alpha$ an arbitrary Grassmann parameter.
We will show  that this symmetry has far reaching consequences for the properties of the fermionic system. 
The non-existence of the Anderson-Higgs effect can be traced to a non-renormalization resulting from
this symmetry.
Note that this worldline supersymmetry does not imply a supersymmetric system in the target space in which
the particle is embedded, which represents just a bosonic system; it acts as a short-hand to capture the particle statistics in the target space.

We now turn our attention to the study of $N$ spin-$\frac{1}{2}$ fermions, which follows
essentially the same logic as for bosons, except that now we must deal with the action given in
\eqref{susy-free} and \eqref{susy-int}, which as already pointed out manifests a worldline supersymmetry.
In the presence of many particles, the worldline can wind many times around the temporal direction, and the functional integral over the Grassman fields we introduced serves to keep track of the exchange
statistics,
\begin{equation} \label{sign} 
  \int{\cal D} \theta \exp\left[ {\frac{i}{4}
  \int_0^{w \beta}d\tau\theta_j\dot{\theta}^j}\right] \:\propto\:
  (-1)^{w-1}
  \ ,
\end{equation}
providing a minus sign for an even winding (corresponding to an odd permutation), and a plus sign for odd winding (corresponding to an even permutation), and the proportionality constant counting the
number of fermionic degrees of freedom.
This sign agrees with the $(-1)^{w-1}$ factor in the grand-canonical formalism \eqref{free}.

Proceeding as we did earlier for bosons, we set the electric field to zero, pick out the long windings and expand the interaction term, and we get for the effective action  the correction 
\begin{equation}
  \Delta \Gamma(w\beta) \:=\:
  \frac{e^2}{2} \Big< \int d\tau_1 \int d\tau_2 \
 \left\lbrace  {\dot x}_i(\tau_1) A^i(x(\tau_1)) \
  {\dot x}_j(\tau_2) A^j(x(\tau_2))-\frac{1}{(4M)^2}F_{ij}(\tau_1)\theta^i(\tau_1)\theta^j(\tau_1) F_{kl}(\tau_2)\theta^k(\tau_2)\theta^l(\tau_2)\right\rbrace 
  \Big>_0 \ ,
\end{equation}
Here in addition to the bosonic Green's function, we also need the fermionic
contribution
\begin{equation}
  2\delta^{ij}G^f_w(\tau_1,\tau_2)\equiv\left\langle \theta^i(\tau_1)\theta^j(\tau_2)\right\rangle_{N,0} \ ,
\end{equation}
which is calculated to be
\begin{equation}
  G^f_w(\tau_1,\tau_2)=\frac{1}{2}\theta(\tau_1-\tau_2) \ ,
\end{equation}
with $\theta(\tau)$ the step function. Carrying out the steps as before, we get the same result for the effective actions as in \eqref{mast}, with the observation that we need to add a fermionic contribution to $\Omega_{ij}(k)$, which shifts the term $(-\frac{1}{2}+\frac{\tau}{N\beta})^2$ to $(-\frac{1}{2}+\frac{\tau}{N\beta})^2-\frac{1}{4}$, that is 
\begin{equation}
\label{omeferm} 
   \Omega_{ij}(k) = (k^2\delta^{ij}-k^i k^j)
\int_0^{w\beta}  d\tau \left( (-\frac{1}{2}+\frac{\tau}{w\beta})^2-\frac{1}{4}\right) 
  e^{\frac{k^2}{M}(-\frac{\tau}{2}+\frac{\tau^2}{2w\beta})} \ .
\end{equation}
and the new addition will make a critical impact.

The qualitative picture remains the same for $w$ finite. There is still the vacuum polarization effect.
However, in the limit $w\to \infty$, the picture changes completely. The integral appearing in
$\Omega_{ij}(k)$ \eqref{omeferm}, becomes
\begin{equation}
  \int_0^{w\beta}d\tau \left[ (-\frac{1}{2}+\frac{\tau}{w\beta})^2 -\frac{1}{4}\right]
  e^{\frac{k^2}{M}(-\frac{\tau}{2}+\frac{\tau^2}{2w\beta})}=
  \frac{M}{k^2} \left( 1-\frac{\sqrt{\pi}}{2}\frac{({\tilde w}+2)}{\sqrt{\tilde w}}e^{-\tilde w/4}
  {\rm Erfi}[\sqrt{{\tilde w}}/2] \right) \ ,
\label{error} 
\end{equation}
with ${\tilde w}=w\beta\frac{k^2}{2M}$, and the imaginary error function given by ${\rm Erfi}[x]=\frac{2}{\sqrt{\pi}}\int_0^x e^{t^2}dt$.
One can see that the function (\ref{error}) vanishes in the limit $w\to\infty$,
thus \begin{equation}
\lim_{w\to\infty}\Omega_{ij}(k)=0,
\end{equation}
a result that can also arrived at by making a saddle point expansion of the left-hand side.
That is to say that the contribution to the effective action arising from the fermionic part cancels precisely that of the bosonic part in the limit of large $N$. In the Grassmannian language, it is the worldline supersymmetry between the bosonic coordinate $x$ and the fermionic coordinate $\theta$ that destroys the Anderson-Higgs or Meissner
effect. We note again that the Grassmann fields simply encapsulate the fermion signs, and it is these signs
which transform the behaviour in the case of fermions in an essential way.


\section{Conclusions}

Below some finite critical temperature, infinitely long windings proliferate
in both bosonic and fermionic systems. For the former this drives Bose-Einstein condensation, while for the
latter it occurs at the Fermi temperature $T_F$.
Owing to the statistics of the particles involved though, the long windings generate vastly different
physics. For the Bose system, it gives rise to superfluidity for neutral systems and superconductivity for
charged ones.
Both are consequences of spontaneous symmetry breaking, breaking a global symmetry in the
neutral superfluid and a gauge symmetry for the charged superconductor (seen as the
Anderson-Higgs mechanism, and responsible for the Meissner effect).

We have attacked the question of whether fermions can drive spontaneous symmetry breaking of a local nature
with the tools of the signful path integral.
We managed the fermion signs by introducing a new Grassmannian coordinate, leading to a supersymmetric
worldline theory. It is supersymmetry then that eliminates the Meissner effect for a gas
of charged fermions. The question still remains whether one can find an order parameter
for the phase transition involving fermions, even in the free case, and how to understand the sharpness
of the Fermi surface.



\section{Aknowledgements}

This research was supported by the �Nederlandse organisatie voor Wetenschappelijk Onderzoek� (NWO)
and by the �Stichting voor Fundamenteel Onderzoek der Materie� (FOM). 


\bibliographystyle{apsrev}
\bibliography{higgsnew}

\end{document}